\documentclass[prepring,journal]{IEEEtran}
\ifCLASSINFOpdf
\else
\fi
\usepackage{url} 
\usepackage{lineno}
\usepackage{soul}
\usepackage{xcolor}
\usepackage{array}
\usepackage[english]{babel} 
\usepackage{adjustbox}
\usepackage{amssymb}
\usepackage{amsmath}
\usepackage{txfonts}
\usepackage{mathdots}
\usepackage[classicReIm]{kpfonts}
\usepackage{graphicx} 
\usepackage{algorithm2e}
\usepackage{booktabs}
\usepackage{sidecap}    
\usepackage[utf8]{inputenc}
\usepackage{ragged2e}
\immediate\write18{texcount -tex -sum  \jobname.tex > \jobname.wordcount.tex}
\usepackage{sidecap}    

\usepackage{subfig}

\begin{document}
%
\title{Long-term temporal-scales of hydrosphere changes observed by GPS over Europe: a comparison with GRACE and ENSO}
%
%
%

\author{Ga\"el Kermarrec         \and
        Anna Klos \and Artur Lenczuk \and Janusz Bogusz 
}

\author{Ga\"el Kermarrec,~\IEEEmembership{}
         Anna Klos,~\IEEEmembership{}
         Artur Lenczuk,~\IEEEmembership{}
        and~Janusz Bogusz ~\IEEEmembership{}

\thanks{Manuscript received ; revised .
We thank H. Dobslaw and A. Eicker for discussions.
GK is supported by the Deutsche Forschungsgemeinschaft under the project KE2453/2-1. AK, AL and JB are financed by the National Science Centre, Poland, grant no. UMO-2022/45/B/ST10/0033. 
(Corresponding authors: Ga\"el Kermarrec and Anna Klos.)}

\thanks{G. Kermarrec is with the Institute for Meteorology and Climatology, Leibniz Universit\"at Hanover, Hanover, Germany,
e-mail: kermarrec@meteo.uni-hannover.de.
A. Klos, A. Lenczuk  and J. Bogusz are with the Military University of Technology, Warsaw, Poland, e-mail: anna.klos@wat.edu.pl}}
%
%

\markboth{IEEE Geoscience and Remote Sensing Letters,~Vol., No., }%
{Shell \MakeLowercase{\textit{et al.}}: Bare Demo of IEEEtran.cls for Journals}
%



\maketitle

\begin{abstract}

Hydrogeodesy can benefit greatly from the use of Global Positioning System (GPS) displacements to analyse local changes in the hydrosphere, which the commonly used Gravity Recovery and Climate Experiment (GRACE) mission is unable to provide due to coarse spatial resolution. Hydrosphere changes recorded by GPS are unfortunately hidden among the other signals to which the system is also sensitive so that the sensitivity of GPS to changes in the hydrosphere on temporal-scales from pluri-annual to decadal is questionable. We focus on hydrosphere signatures present on these long-term temporal-scales as observed by GPS through the vertical displacement time series (DTS) of 122 permanent stations over Europe and compare them to the DTS derived from GRACE for GPS locations. Our methodology is based on the weighted Savitzky-Golay (S-G) filter, an underestimated filter in the field of geodetic time series analysis. We show that the correspondence between GPS and GRACE on long-term temporal-scales is generally strong, but decreases for coastal regions and regions where the coarse gridding of GRACE does not capture local hydrosphere effects. Further, the negative correlation with El Niño Southern Oscillations (ENSO) is confirmed for Europe.

\end{abstract}


\IEEEpeerreviewmaketitle

\section{Introduction}

\IEEEPARstart{A}ccelerating climate change is causing Europe to face more frequent sudden floods or prolonged droughts making an improved understanding of hydrospheric changes mandatory \cite{Boergens2020,Lehmkuhletal2022}. Continuous monitoring of Total Water Storage (TWS \cite{Rodell2004}) is the main task of hydrogeodesy, a new field using geodetic observations for this purpose. TWS changes have been estimated since 2002 based on observations from the GRACE mission \cite{Landerer2020,Humphrey2023,Whiteetal2023}, but their low spatial resolution of approximately 300 km introduces considerable interpretive limitations in analyzing local changes. Thus, GRACE estimates of TWS changes are reliable only at the basin-scale level for Europe, and only in the eastern part of the continent due to the irregular coastline, leakage effect between land and oceans and high inland water intrusion \cite{Bian}. The use of GPS DTS are an alternative, boosted since 2000 \cite{Blewittetal2001}. Subsequent analyses have shown the compatibility of the DTS with the displacements derived by GRACE for GPS locations \cite{Davisetal2004, VanDametal2007}. \cite{Argusetal2014} were the first to use GPS displacements to estimate TWS through inversion procedure and show that GPS displacements yield hydrosphere changes with spatial resolution as high as a few tenths of a degree \cite{Argus2017, White2022}. 
Permanent GPS stations have a worldwide spatial coverage, satisfactory for most regions \cite{Blewitt2018}. The GPS DTS are provided with daily temporal resolution, which can greatly facilitate interpretations of hydrosphere events. 

Sensitivity of GPS DTS to seasonal changes in the hydrosphere is known to  be pronounced for eastern Europe, Misissippi basin or Amazon area \cite{Becker2011}. For other areas, a difference in seasonal phase values between GPS-observed and GRACE-derived DTS is often observed: An overlap with the other contributors within GPS DTS (draconitic signals, thermal expansion of ground or environmental loading effects \cite{Michel2021}) are potential reasons. GPS DTS are believed still not to be fully trusted for determining hydrosphere signatures at long-term temporal-scales, from pluri-annual to decadal. In this contribution, we propose to study these temporal-scales for GRACE-derived and GPS-observed DTS. Several studies have shown that these signatures correspond well between both DTS, but mainly focused on regions that experienced extreme hydrosphere changes (Australia and the Amazon basin \cite{Dill2013}). No one has made this comparison over Europe, a region where the number of long-term hydrosphere events is increasing due to climate change \cite{Pfefferetal2023}: 4 long-term droughts were recorded from 2002 to now in Europe. While GRACE can provide their approximate location, exact insight can only be given by GPS, especially for regions where GRACE estimates may not be accurate. This is of colossal importance when analyzing groundwater loss due to prolonged droughts, which can leave a densely populated area of Europe without access to potable water.  

To answer the question "Can the long-term temporal-scale of GPS DTS be trusted for hydrosphere study over Europe?", we filter the station-specific high frequency signal (around the annual frequency and higher) from GRACE-derived and GPS-observed DTS. Our focus is on a fine-tuned weighted S-G filter \cite{Savitzky1964}, which we compare with a moving-average smoothing filter \cite{White2022}, and the Whittaker Henderson (WH) smoothing filter, sometimes called a perfect smoother \cite{eilers2003}. Although a classical method, the S-G filter is robust, computational efficient, and, most importantly, preserves the shape (peaks and features) of the filtered time series, an important property for our study. \cite{Liu} used the S-G filter for seismic signals, \cite{Roy} for self-potential anomaly, or \cite{Gou} in a pre-processing step combined with deep learning algorithms for length of day time series. This powerful filter deserves more attention within a geodetic context. Our new application shows that a strong anticorrelation exists between filtered GPS-observed DTS and ENSO for the most part of Europe, as previously highlighted using GRACE over East China \cite{HE2020124475}, and is a promising result.

In Section~\ref{Sect:description}, we describe the datasets utilized. Section~\ref{sect:methodology} presents the S-G filter. In Section~\ref{sect:results}, we comment on the differences and similarities between the high-pass filtered DTS and focus on specific stations. 

\section{Datasets} \label{Sect:description}

\subsection{GPS-observed DTS}

We use daily vertical DTS processed by the Nevada Geodetic Laboratory (NGL, \cite{Blewitt2018}) for the period coinciding with GRACE, i.e. 2002-2018. The NGL solution has the advantage of being updated automatically in near real time, which is very beneficial for studying unexpected hydrosphere effects. It includes more than 20,000 permanent stations making regional analyses possible. The potential of the NGL solution is often underestimated in the geodetic community, but recent studies have shown a fairly high correspondence between the NGL GPS DTS and DTS derived by GRACE \cite{Knowles2020, Klosetal2023}. We select DTS for 122 GPS European stations for which the DTS are believed to be of the highest quality within the geodetic community. GPS observations were processed in a Precise Point Positioning mode in a GipsyX Version 1.0 software using final Jet Propulsion Laboratory (JPL) Repro-3 products, such as satellite orbits and clocks \cite{Blewitt2018}. Higher-order ionospheric effects were accounted for using IONEX data from JPL, while tropospheric effects were modeled using the VMF1 function. Phase center offsets were applied using the IGS model, solid Earth and pole tides were removed using the IERS Conventions (2010), and ocean tidal loading modeled using the FES2004 tidal model. The NGL GPS DTS are expressed in a center-of-mass reference frame and are pre-processed to remove outliers and offsets using three times the interquartile range rule and available databases of offsets, supported with our manual inspection, on a station-by-station basis. 

Non-tidal environmental loading effects which are not commonly applied at the processing level contribute to the GPS DTS at different temporal-scales, depending on the location \cite{Memin2022,Klos2021}. We remove the effects of non-tidal atmospheric loading (NTAL), non-tidal oceanic loading (NTOL) and a response to barystatic sea-level variability (SLEL) using predictions of vertical DTS provided by the Earth System Modelling Group of Deutsches GeoForschungsZetrum, as in \cite{Dill2013}. Predictions of vertical DTS resulting from NTAL and NTOL are resampled from three hours to daily samples, which agree with the DTS observed by GPS. Predictions of vertical DTS resulting from SLEL are left in their original 24-hour samples. The thermal expansion of bedrocks and monuments was neglected as the effects often do not exceed a tenth of the amplitude present in the vertical DTS observed by GPS \cite{Xu2017}. All DTS predicted by a model were interpolated from their original $0.5^{\circ}$ by $0.5^{\circ}$ grids into the location of GPS stations and removed from the GPS DTS. Finally, GPS DTS were detrended using the least-squares method and resampled from daily into monthly samples, for a comparison to GRACE-derived DTS. Since all known effects contributing to the long-term temporal-scale of GPS DTS are removed, we assume the remaining signals originate from hydrosphere loading.

\subsection{GRACE-derived DTS}

We use the GRACE-derived TWS values available in the form of iterated global mascon solution provided by the Center for Space Research in Austin as the latest release-06 (RL06) \cite{Save}. The mascons were estimated using Level-1 observations, with several effects removed during processing stage: static gravity field removed using GGM05C \cite{Ries2016}, atmospheric and oceanic effects removed using the AOD1B RL06 based on the European Centre for Medium-Range Weather Forecast and Max-Planck-Institute for Meteorology Ocean Model models \cite{Dobslaw2017}, polar tides of the solid Earth and ocean removed using the GOT4.8 model with the Self-Consistent EQuilibrium model up to degree and order 180 \cite{Ray1994}. Degree-1 coefficients are replaced following \cite{Sun2016} using the estimates in Technical Note 13 (TN13). The C20 coefficients are replaced for GRACE by Satellite Laser Ranging estimates using TN14 \cite{Loomis2019}. The GIA effect was removed using the ICE6G\_D \cite{Peltier2018}. TWS is provided as anomalies relative to 2004.0000 - 2009.999 mean baseline \cite{Save}. We then transform monthly gridded TWS values into monthly vertical DTS for locations of 122 GPS stations and remove trends using the least-squares method. Finally, we selected epochs common to GPS-observed and GRACE-derived DTS. We disregard the period after 2017 intentionally to avoid having to interpolate the gap between GRACE and GRACE Follow-On, which may affect our interpretation.


\section{Methodology}\label{sect:methodology}

\subsection{S-G filter and its tuning}

In this contribution, we focus on filtering out the high frequencies from GRACE-derived and GPS-observed DTS using the S-G filter~\cite{Savitzky1964} with a Kaiser window (KW) weighting to limit boundary effects \cite{Schmid}. We extract the long-term temporal scales (strictly \(>1\) year) and not simply perform a traditional smoothing which would eliminate the very high frequency (\(<1\) year) \textit{only}.
The S-G filter is based on a least-squares fit of a small set of consecutive data points to a polynomial. The calculated central point of the fitted polynomial curve is taken as the new smoothed data point in a second iteration. Two parameters have to be tuned to reach the low-pass filtering effect at a given cut-off frequency: (i) the polynomial order $d$, set to the typical value of 3 to avoid overfitting, and (ii) the half-width of the smoothing window $m$. A large value of $m$ will produce a smoother DTS at the expense of flattening the peaks or pseudo-periodic components as illustrated in Fig.\ref{fig:SFER_SGwindows} (a) with the GRACE-derived DTS of station SFER in Spain. A small $m$ preserves the high frequency and periodical (yearly) components. We fix $m$ based on the 3dB cutoff frequency $fc$, which is given empirically by $fc=\frac{d+1}{3.2m-4.6}$ \cite{Schafer2011}. Considering that the monthly GRACE-derived and GPS-observed DTS have strong energy in the filter passband, we adopt a value of $m=61$, i.e., a cutoff under 1.3 year approximately. This way, we strictly prevent ourselves from filtering frequencies higher than one year.
An adjustment of the filter is required near the sample boundaries when the window extends beyond the beginning or the end of the input vector, as illustrated in Fig.\ref{fig:SFER_SGwindowsKF3} (b). We use a KW to weight the time series as proposed in \cite{Schmid}. A value of 3 was found optimal for GPS-observed DTS. 

\subsection{Comparison to other filters}

\begin{figure}          
\centering
\begin{tabular}{cc}
    {\includegraphics[width=0.20\textwidth]{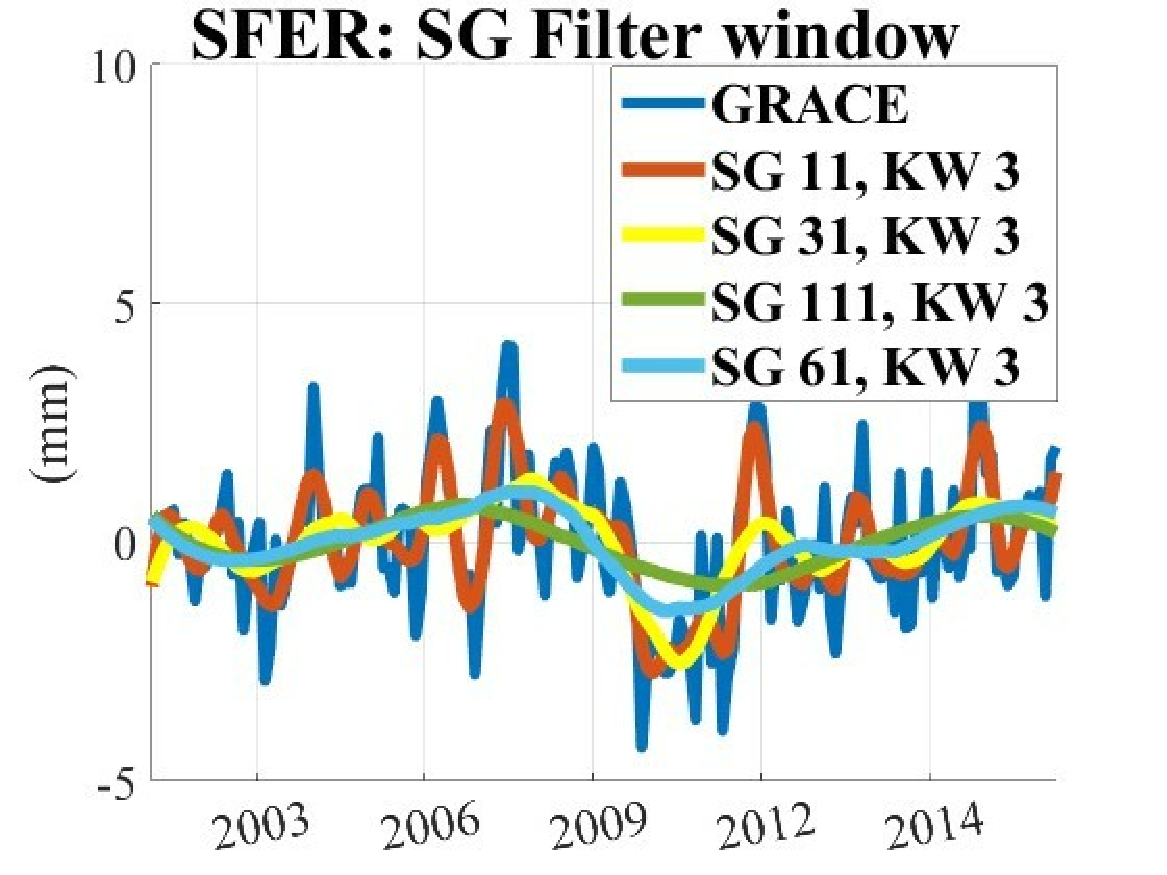} } 
&
      {\includegraphics[width=0.20\textwidth]{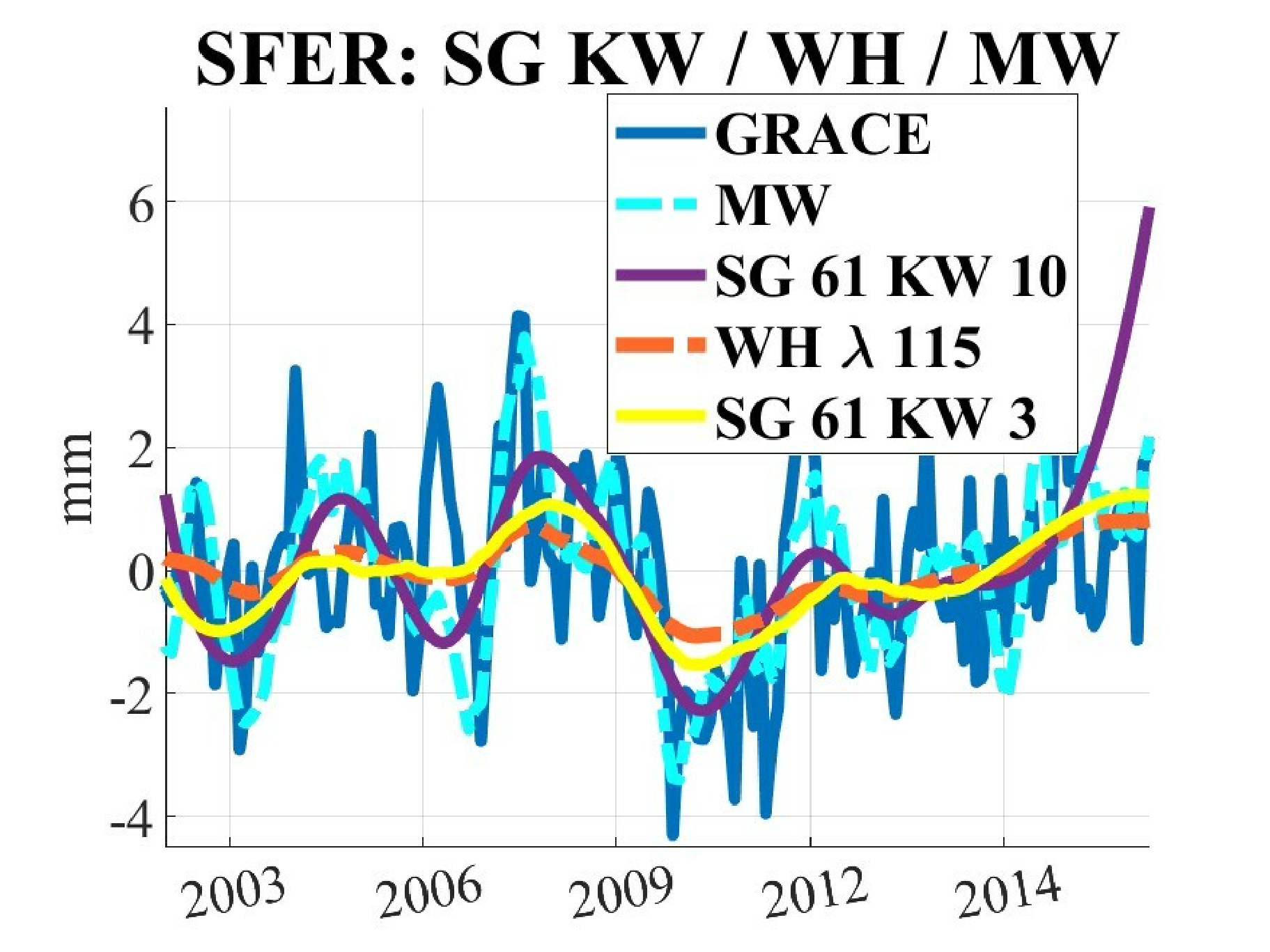} }
\\
(a)&(b)
\end{tabular}
\caption{Station SFER (Spain). (a): S-G filtered detrended GRACE-derived DTS for different values of $m$. (b): same using a KW, a moving average and a WH filter.}
\label{fig:SFER_SGwindows}
\label{fig:SFER_SGwindowsKF3}

\end{figure}

We illustrate the difference between the S-G filter and a traditional moving-average smoothing approach in Fig.\ref{fig:SFER_SGwindowsKF3} (b). With this method, only high frequencies have been suppressed (dot cyan), without extracting the long-term temporal-scales specifically, which are the topic of this contribution. We, thus, will not investigate the results using a smoothing filter further. 

The main advantage of the S-G filter over similar filters (binomial/Gaussian or Legendre filter \cite{Persson2003}) comes from its flat passband with low attenuation in its stopband which avoids masking or downscaling the long-term temporal-scale. An alternative to the S-G filter is the a non-ﬁnite
impulse response (FIR) method called the WH filter. This approach is based on spline smoothing (\cite{Schmid2022} and the references inside). The cut-off frequency can be tuned similarly to the S-G filter to lowpass a specific frequency band. We chose a similar set-up as for the S-G filter with a penalty factor $\lambda=115$ and a degree of 3 for the derivative. The result of the filtering is presented in Fig.\ref{fig:SFER_SGwindowsKF3} (b): No boundary effect occurs, but the peak amplitudes are weaker than for the S-G filter in the middle of the TS. 
We are strong believers that the peak-preservation is the major advantage of the S-G filter to analyse hydrosphere DTS and identify years of higher variability with trustworthiness. For the sake of completeness, we performed the same investigations as in Sect.\ref{sect:results} with the WH filter by comparing WH filtered GRACE-derived and GPS-observed DTS. If the values differed slightly compared to the S-G filter (not more than 5\%), the relative comparison showed similar pattern and areas of high or low correspondence for both methods. For the aforementioned reasons, we prefer the highly efficient S-G filter for our new application. The fact that it is an old technique does not put it out of play compared to machine learning approach. The S-G filter is constantly improved with, e.g., colored noise models \cite{Kennedy}, and has various and recent applications \cite{Gajbhiye}.

\section{Results}\label{sect:results}

We compare the low frequency (LF) components of the GRACE-derived and GPS-observed DTS after and before subtraction of loading products as described before. We use the Pearson correlation coefficient (PCC) to quantify similarities between DTS \cite{FANG2021441}. The results are presented Europe-wise for the 122 stations under consideration. We illustrate the methodology using specific time series additionally.

\subsection{Europe-wise comparison}

\subsubsection{With and without loading products}
\begin{figure}          
\centering
\begin{tabular}{cc}
    {\includegraphics[width=0.21\textwidth]{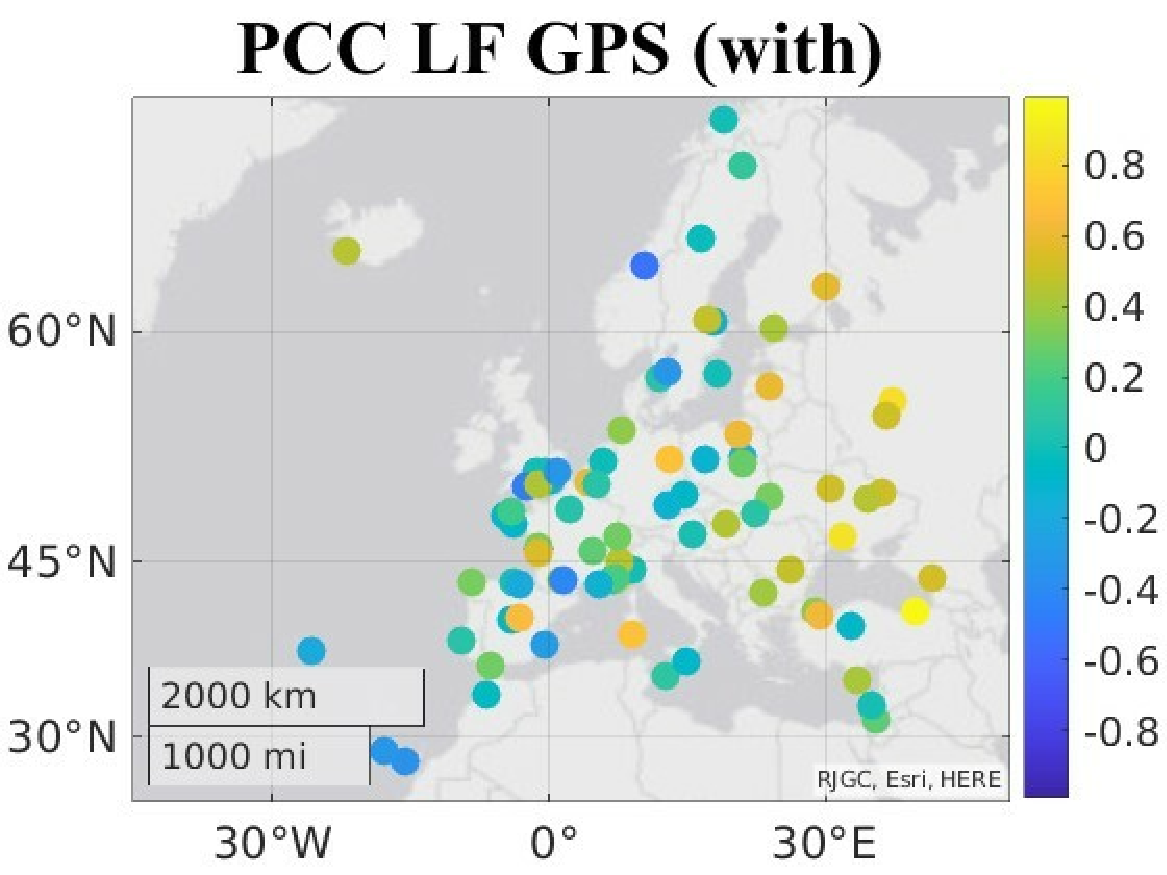} } 
&
      {\includegraphics[width=0.21\textwidth]{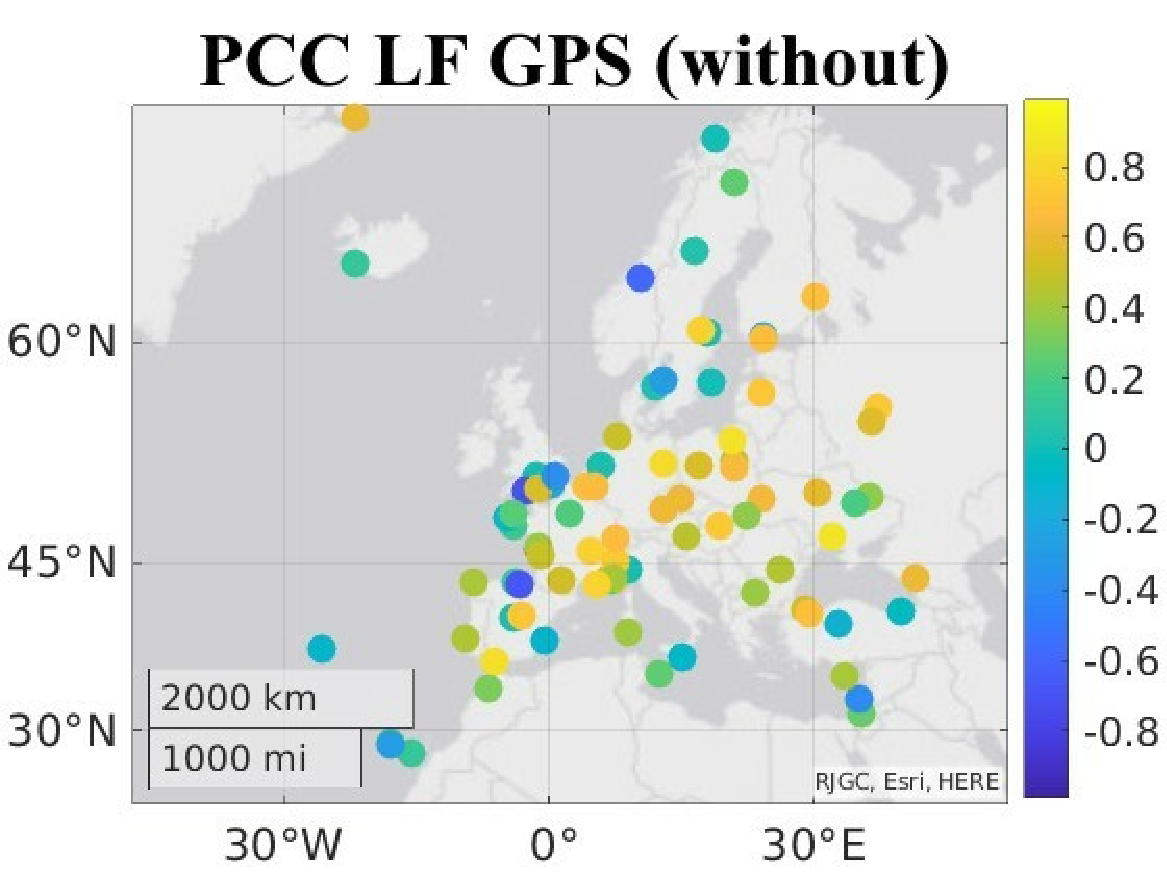} }\\
      (a)&(b)
\end{tabular}
\caption{PCC between LF of GRACE and GPS DTS (a) before and (b) after correction with environmental loading products.} \label{fig:NGLglobal}
\end{figure}

A visual comparison between Fig.\ref{fig:NGLglobal} (a) and (b) highlights the impact of environmental loading products on the LF component of GPS-observed DTS. After subtraction in  Fig.\ref{fig:NGLglobal} (b), the PCC increases compared to (a) (yellow dominates - PCC$>$0.6). We found a mean value of 0.45 after subtraction. The similarity at LF is particularly high in the center of Europe whereas many stations in Scandinavia, near the Atlantic ocean and the Mediterranean sea have either a PCC close to 0 or are anticorrelated (dark green-dark blue). 
This effect is present at nearby stations, which gives weight to the interpretation that it should be related to hydrosphere effects that are not captured by GRACE-derived DTS. The high variability in Scandinavia and Svalbard corresponds to a region for which GRACE may not be reliable due to an impact of Baltic Sea and Arctic Ocean. Here GPS-observed DTS are more reliable in catching the LF of hydrosphere mass loading. The same holds true for stations near the Atlantic ocean where the coarse gridding of GRACE is unfavorable. 
We mention that residuals from NTOL for stations close to the sea may be present in the DTS observed by GPS, but at a frequency which is most probably eliminated with our filtering strategy.

When predictions of displacements of environmental loadings are eliminated (Fig.\ref{fig:NGLglobal} (a)), the PCC between the LF components of GRACE-derived and GPS-observed DTS is lower than before for nearly all stations under consideration (mean 0.15). The methodology consisting of subtraction products, even if these latter may not be perfect, increases the similarities between GRACE-derived and GPS-observed DTS at LF significantly.

\subsection{Specific time series and comparison to ENSO}\label{sect:timeseries}

\begin{figure}          
\centering
\begin{tabular}{cc}
    {\includegraphics[width=0.21\textwidth]{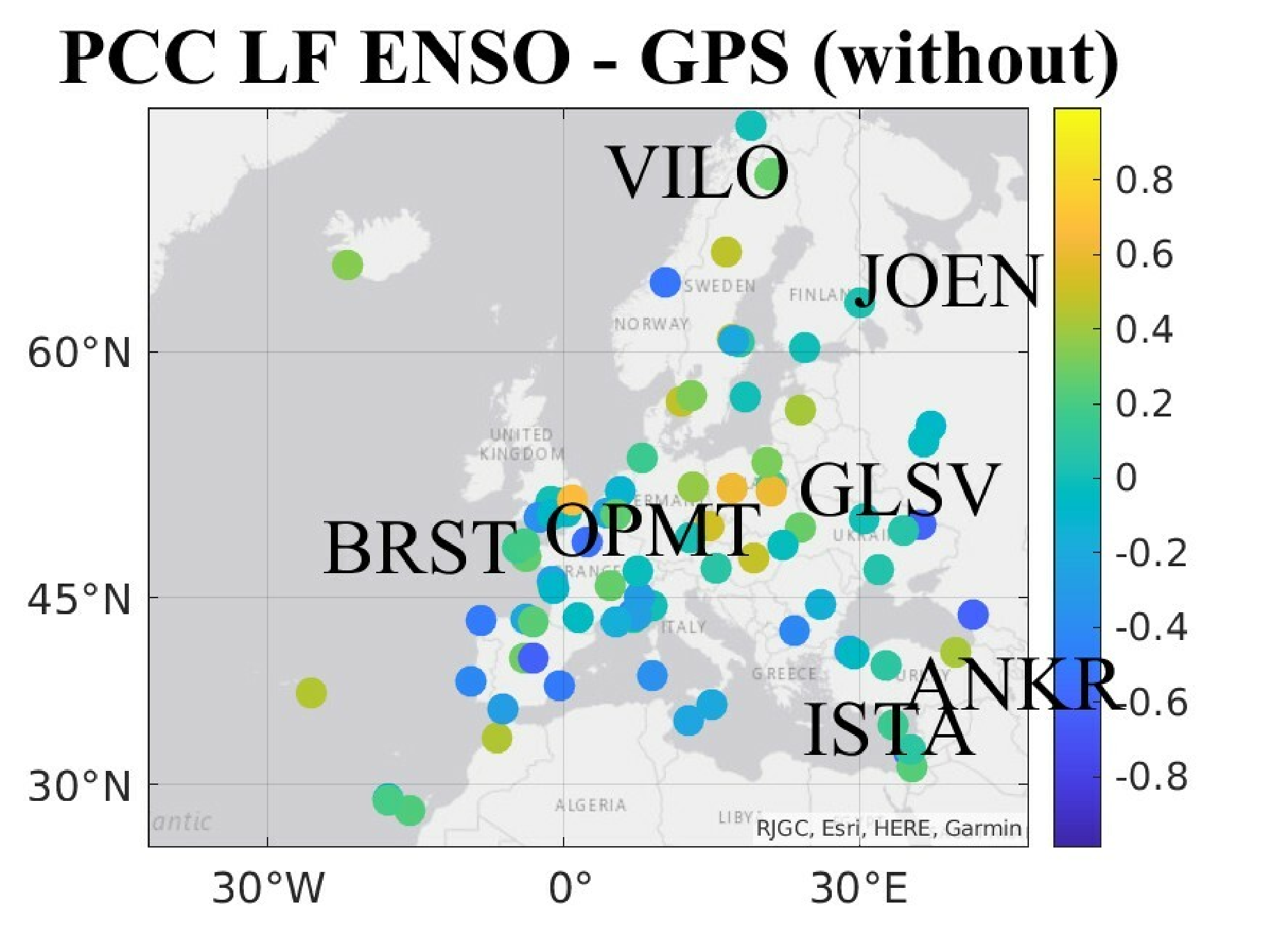}} 
&
      {\includegraphics[width=0.21\textwidth]{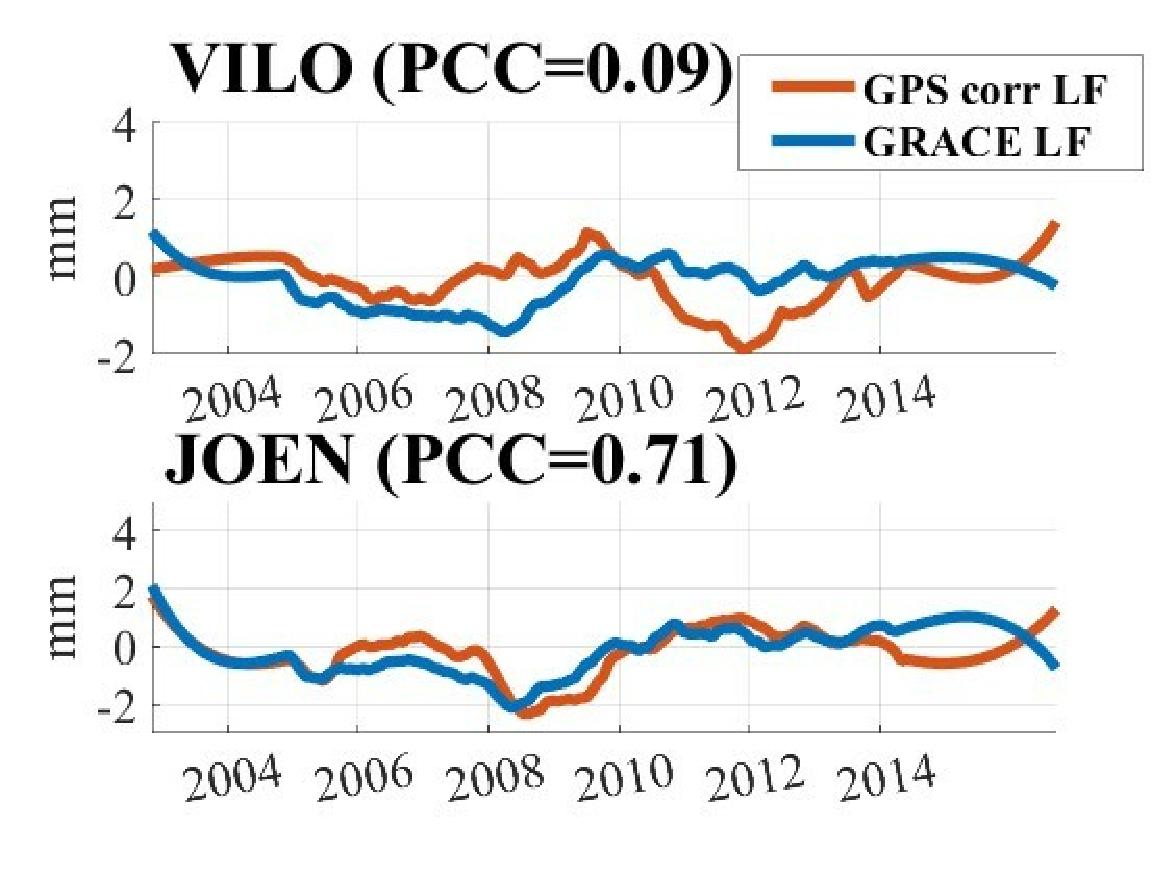} }\\
      (a)&(b)\\
      
          {\includegraphics[width=0.21\textwidth]{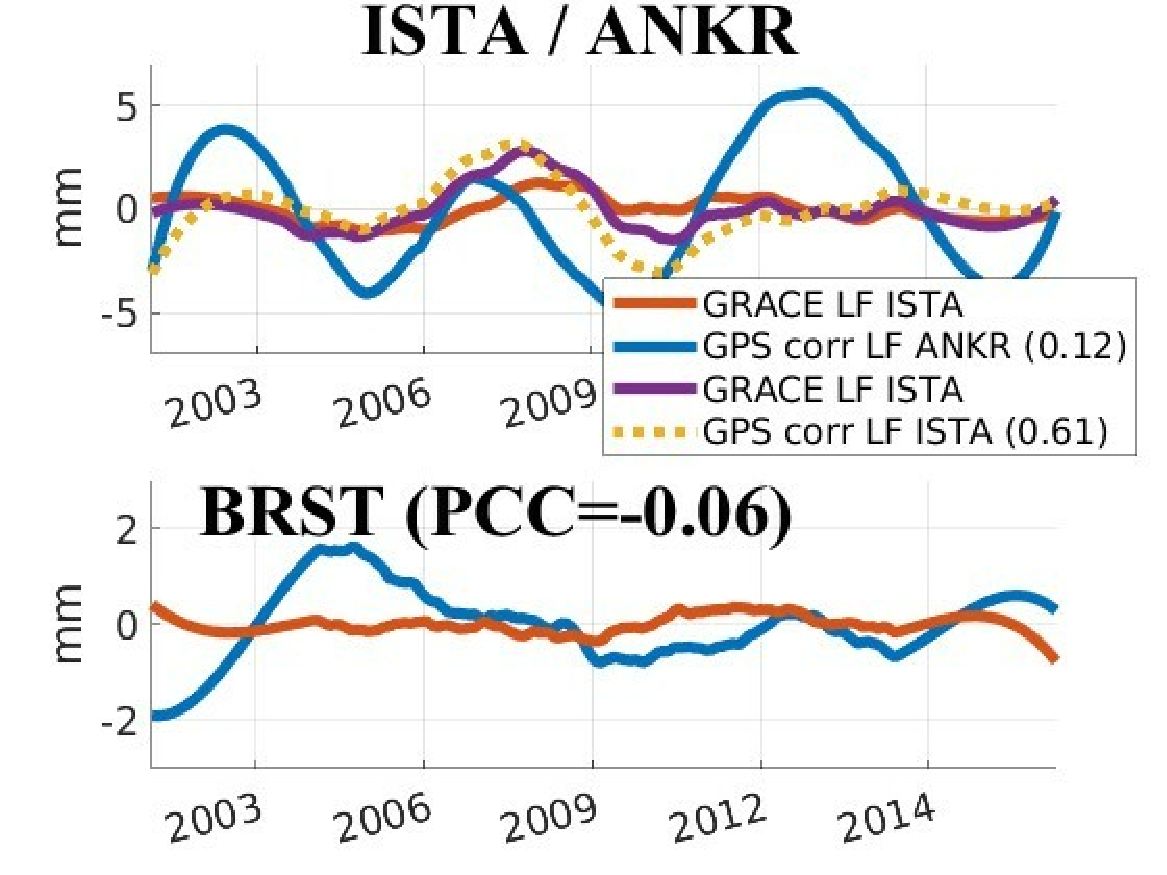} } 
&
      {\includegraphics[width=0.21\textwidth]{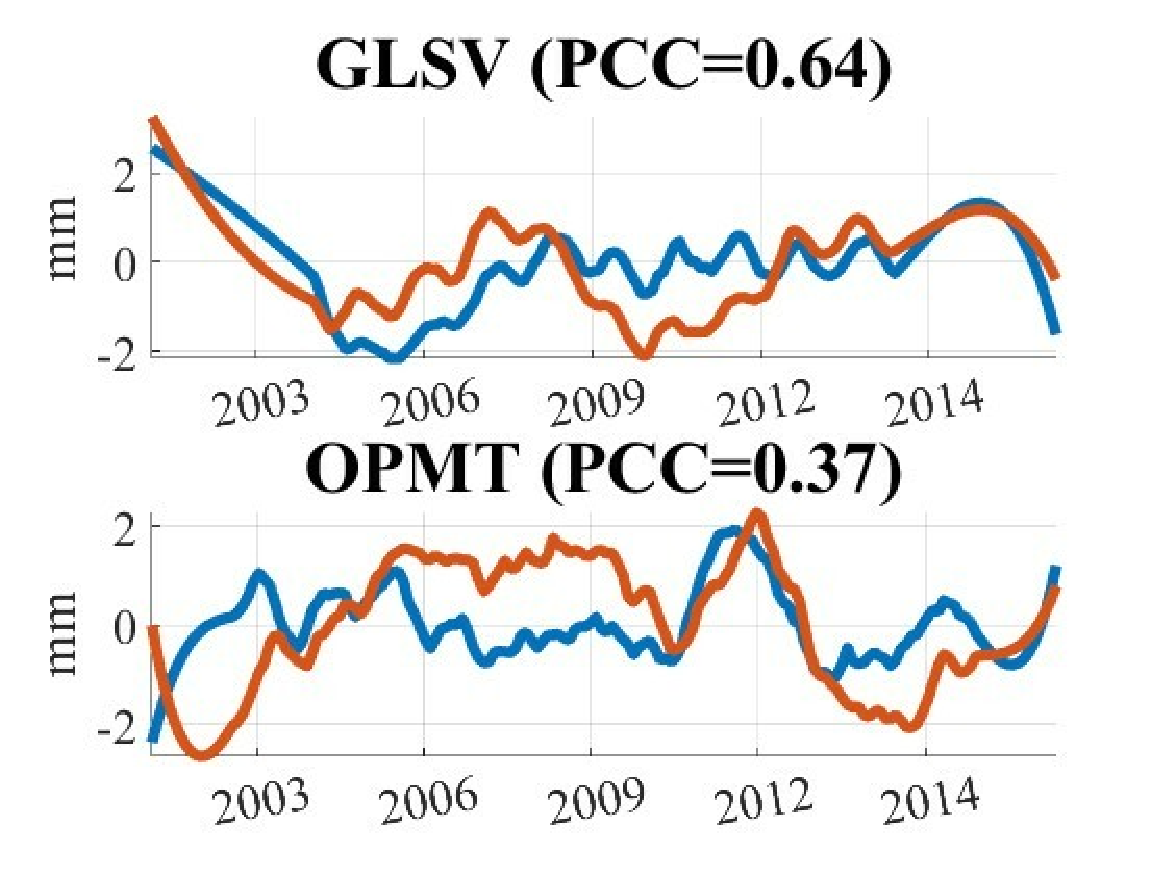} }\\
      (c)&(d)
\end{tabular}
\caption{(a) localisation of the stations, (b) VILO and JOEN, (c) ANKR, ISTA and BRST, (d) GLSV and OPMT. Red: GPS DTS, blue: GRACE-derived DTS.}

\label{fig:timeseries}

\end{figure}

We haven chosen six specific stations located as in Fig.\ref{fig:timeseries} (a) for further investigations.
JOEN (b) is situated in Scandinavia (Finland) and is associated with a high PCC ($>$0.7). BRST (c) is a GPS station in Brittany (France) for which the PCC is close to 0: GRACE seems to underestimate the LF amplitude (coarse gridding and low trustworthiness in coastal regions). The station VIL0 (Sweden) has also a low PCC (0.2). Here we note that the LF amplitude from GRACE slightly increased between 2009 and 2011 whereas GPS-observed DTS decreased. This effect (not shown) is also visible for TRDS which is spatially close to VIL0. The GPS-observed DTS is time-shifted with respect to GRACE-derived DTS (see year 2008 in Fig.\ref{fig:timeseries} (b)). Here, GPS should record a slightly different hydrosphere effect than GRACE, which sensibility to hydrosphere changes is low in that region.
We consider further the station ANKR and ISTA (c) in Turkey for which the PCC is small. The discrepancy between GRACE and GPS may be related to very local tectonic activity (ANKR) and not to hydrosphere.
We further analyse the LF component from two stations: GLSV (Kiev, Ukraine), and OPMT (Paris, France). 
For the GLSV station (d), GPS catches a specific effect between 2010-2011. Indeed, the decrease in DTS is most likely due to the large amount of snowfall and the accumulation of ice on the dam on the Dnieper River as a result of which large amounts of water had to be drained to prevent the risk of springtime flooding of neighboring areas (source: https://earthobservatory.nasa.gov/). For the OPMT station (d), GPS captures variations probably indicating the subsidence of Parisian basin \cite{Prijac}, but not GRACE (flat blue line).

We have performed additionally a comparison with ENSO, from which we filtered the same long-term temporal-scales as for GPS-observed DTS. The results are presented in Fig.\ref{fig:timeseries} (a). A strong anticorrelation (approximately $4-6$ months delay) is generally found except in the middle of Europe. This effect could be partially explained by specific regional climate variabilities \cite{HE2020124475}. Further investigations are mandatory.

\section{Conclusions and outlook}
We developed a filtering strategy to extract the long-term temporal-scales from vertical DTS observed by GPS and derived from GRACE. We found a high correlation between the GRACE-derived and GPS-observed DTS after correction with environmental loading products. For specific regions and stations in Europe, a discrepancy may be caused by the coarse gridding of GRACE and its low capacity of catching effects in coastal regions. Stations with tectonic activity show a lower correlation. The GPS-observed DTS are anticorrelated with ENSO in most part of Europe. This effect remains to be deeper investigated but is a promising application. NGL-based GPS DTS have the advantages of being available for a huge number of stations worldwide and updated regularly with short latency. We showed their potential for inclusion into hydrogeodetic assessments.

\ifCLASSOPTIONcaptionsoff
  \newpage
\fi



\bibliographystyle{IEEEtran}
\bibliography{revision_GRSL}
*\section{comment}
ieeedoi:10.1109/LGRS.2023.3345540
\end{document}